\documentclass[aps,prl,reprint,superscriptaddress]{revtex4-1}
\usepackage{amsfonts,amsmath,amssymb,mathrsfs,dsfont,yfonts,bbm,changepage}
\usepackage[dvipsnames]{xcolor}
\usepackage{mathtools,nccmath}

\newcommand{\be}{\begin{eqnarray}}
\newcommand{\ee}{\end{eqnarray}}
\newcommand{\bea}{\begin{eqnarray}}
\newcommand{\eea}{\end{eqnarray}}

\newcommand{\nn}{\nonumber}
\newcommand{\bn}{\begin{enumerate}}
\newcommand{\en}{\end{enumerate}}

\usepackage{tikz,bm}
\usetikzlibrary{arrows}
\usetikzlibrary{patterns}
\usetikzlibrary{decorations.pathmorphing}
\usepackage{xcolor}
\usepackage{amsmath,amssymb,amsthm,mathdots}

\usetikzlibrary{positioning}
\usetikzlibrary{chains}
\usetikzlibrary{arrows, arrows.meta ,fit,decorations.pathreplacing}
\tikzstyle{every picture}+=[remember picture]
\tikzstyle{na} = [baseline]

\usetikzlibrary{arrows, decorations.markings, calc, fadings, decorations.pathreplacing, patterns, decorations.pathmorphing, positioning}
\tikzset{>={Latex[width=1.5mm,length=1.5mm]}}

\definecolor{myblue}{rgb}{0,0,0.5}
\definecolor{myred}{rgb}{0.5,0,0}
\definecolor{myorange}{rgb}{0,0,0.5}
\definecolor{mygreen}{rgb}{0.46,1.39,1}

\definecolor{myyellow}{rgb}{1.0, 1.0, 0.4}
\definecolor{lightgray}{gray}{0.7}
\definecolor{lightblue}{rgb}{0.12, 0.56, 1.0}
\definecolor{lightred}{rgb}{0.99, 0.56, 0.67}

\definecolor{light-gray}{gray}{0.7}

\definecolor{maroon}{rgb}{0.76, 0.13, 0.28}

\tikzset{snake it/.style={decorate, decoration={snake, amplitude=.4mm, segment length=2mm,
                       post length=0mm,pre length=0mm}}}

\newcommand{\udl}[1]{\mathrm{d} #1 \,}

\newcommand{\Gpq}[1]{\Gamma_e\left( #1\right)}

\def\Gd{\Delta}

\def\gs{\sigma}

\begin{document}

\title{Rethinking mirror symmetry as a local duality on fields} 
\author{Chiung Hwang} 
\affiliation{Department of Applied Mathematics and Theoretical Physics, University of Cambridge, Cambridge CB3 0WA, UK}
\author{Sara Pasquetti} 
\affiliation{Dipartimento di Fisica, Università di Milano-Bicocca, I-20126 Milano, Italy}
\affiliation{INFN, Sezione di Milano-Bicocca, I-20126 Milano, Italy}
\author{Matteo Sacchi} 
\affiliation{Dipartimento di Fisica, Università di Milano-Bicocca, I-20126 Milano, Italy}
\affiliation{INFN, Sezione di Milano-Bicocca, I-20126 Milano, Italy}
\affiliation{Mathematical Institute, University of Oxford, Woodstock Road, Oxford, OX2 6GG, United Kingdom}

\date{\today}

\begin{abstract}
We introduce an algorithm to piecewise dualise linear quivers into their mirror dual. The algorithm uses two basic duality moves and the properties of the $S$-wall which can all be derived by iterative applications of Seiberg-like dualities.
\end{abstract}
\maketitle

\section{Introduction}

$3d$ $\mathcal{N}=4$ theories enjoy a mirror duality which relates pairs of dual theories with Higgs and Coulomb branches of the vacuum moduli space exchanged \cite{Intriligator:1996ex}. If we realise these theories on Hanany--Witten brane set-ups in type IIB string theory with D3-branes suspended  between NS5 and D5-branes, mirror symmetry  can be interpreted as the action of $S$-duality on the brane system \cite{Hanany:1996ie}\footnote{In \cite{Hori:1997zj}, 3d mirror is also realised as T-duality between IIA and IIB string theories.}.

It has been argued that $S$-duality can act \emph{locally} on each 5-brane creating an $S$-duality wall on its right and an $S^{-1}$  wall on its left \cite{Gaiotto:2008ak,Gulotta:2011si}
\begin{equation}\label{eq:branedual}
 \overline{ \text{NS5}}\to S^{-1} \text{D5} \,S\,, \qquad  \text{D5}\to S^{-1}  \text{NS5}\, S\,,
\end{equation}
and  the  $S$-wall intersecting $N$ D3-branes is known to correspond to the $T[SU(N)]$ quiver theory  \cite{Gaiotto:2008ak}.

It is  natural to wonder whether this  local $S$-duality action can be understood in field theory as a local action on the quiver.
In this paper we show that this is indeed possible, thus providing a completely field theoretic and algorithmic derivation of mirror symmetry. Specifically, for each element in the relations \eqref{eq:branedual} we can find a field theory counterpart, allowing us to reinterpret \eqref{eq:branedual} as genuine infra-red (IR) dualities in field theory. Such dualities, together with the properties of the $S$-wall, can then be used to systematically dualize a given quiver into its mirror. Crucially, all the basic dualities needed in our algorithm can be derived using more elementary Seiberg-like dualities, that are dualities that are analogues of Seiberg duality \cite{Seiberg:1994pq} in $4d$.

\begin{figure*}[bt]
\makebox[\linewidth][b]{
\includegraphics[]{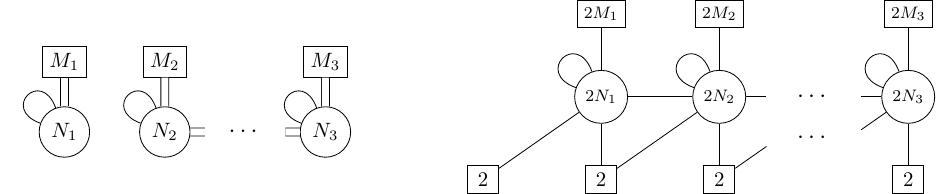} 
}
\caption{\label{rhosigma} On the left the $3d$ $T^\rho_\sigma[SU(N)]$ theory, round nodes denote gauge groups and square boxes flavor groups all of $U(n)$ type. Lines connecting them represent chiral fields. On the right the $4d$   $E_\rho^\sigma[USp(2N)]$ theory (up to singlets), now all nodes denote groups of $USp(2n)$ type. The ranks $N_i$, $M_j$ are given in terms of the partitions $\rho,\sigma$. As explained in \cite{Pasquetti:2019hxf} reducing the $E_\rho^\sigma[USp(2N)]$ to $3d$ and turning two deformations that first break the $USp(2n)$ groups to $U(n)$ and then give mass to the fields of the saw (the $USp(2N_i) \times SU(2)$ bifundamentals) we flow to the $T_\rho^\sigma[SU(N)]$ theory.}
\end{figure*}

Recently in \cite{Hwang:2020wpd} a family of $4d$ $\mathcal{N}=1$ theories called $E_\rho^\sigma[USp(2N)]$, labelled by partitions $\rho,\sigma$ of $N$, were constructed (see  Fig. \ref{rhosigma}).
These theories upon compactification to $3d$ and suitable RG flows reduce to the $T_\rho^\sigma[SU(N)]$ family of unitary gauge linear $3d$  $\mathcal{N}=4$  quivers, first introduced in \cite{Gaiotto:2008ak}\footnote{$T[SU(N)]$ is a special case of $T_\rho^\sigma[SU(N)]$ with $\rho = \sigma = [1^N]$.}. The $E_\rho^\sigma[USp(2N)]$ theories, as their $3d$ relatives, enjoy  mirror symmetry which relates pairs of theories with swapped $\rho$ and $\sigma$ partitions.

One may then ask whether also $4d$ mirror symmetry can be realised  as a local action on the quiver.
We will see that it is indeed possible to define the same algorithm also in $4d$, to locally dualise the fields by means of two basic duality moves, which together with the properties of the $4d$ $S$-wall allow us to go from $E_\rho^\sigma[USp(2N)]$  to its mirror $E_\sigma^\rho[USp(2N)]$. As argued in \cite{Bottini:2021vms}, the $4d$ $S$-wall should be identified with the $FE[USp(2 N)]$ theory \cite{Pasquetti:2019hxf}\footnote{The $FE[USp(2 N)]$ theory is the $E[USp(2 N)]$ theory with some extra gauge singlets.}, which in $3d$  reduces to the $T[SU(N)]$ theory up to gauge singlets. Interestingly, the basic duality moves involved in our algorithm are IR dualities which can be in turn derived by iterative applications of the Intriligator--Pouliot (IP) duality \cite{Intriligator:1995ne} as shown in \cite{Bottini:2021vms}.

Although our discussion here focuses on the $4d$ case, by taking the standard $3d$ limit combined with the suitable Coulomb branch VEVs and real mass deformations, we answer the same question in $3d$, that is we have an algorithm to locally dualise $3d$ $\mathcal{N}=4$ quivers. 

Early attempts to answer the same kind of question in the $3d$ set-up \cite{Assel:2014awa,Gulotta:2011si} reformulated the local $SL(2,\mathbb{Z})$ action at the level of the $\mathbb{S}^3$ partition function without providing the field theory interpretation in terms of applications of genuine IR dualities. 
In the abelian case, the local $S$-duality action can be realised as a piecewise dualisation of a free hypermultiplet into SQED, which was understood as a generalised Fourier transformation of its partition function \cite{Kapustin:1999ha}. Our result is a generalisation of the piecewise dualisation of the 3d abelian mirrors in \cite{Kapustin:1999ha} to the non-abelian case.

One important feature of our dualisation algorithm is the propagation of certain operator VEVs along a quiver via Higgs mechanism, which resembles Hanany--Witten transitions in brane set-ups. Such Higgs mechanism plays an essential role to realise the expected gauge groups in the mirror dual frame.

\section{The 4d $S$-wall}

\begin{figure*}[bt]
\makebox[\linewidth][c]{
\includegraphics[]{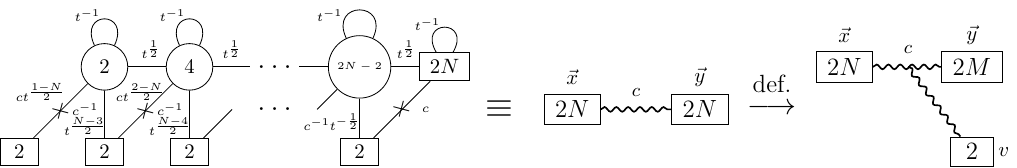} 
}
\caption{\label{fig:fe} The $FE[USp(2N)]$ quiver and its compact representation displaying the manifest and emergent $USp(2N)$ symmetries and the abelian $U(1)_c$ symmetry fugacity. The powers of $t$ and $c$ encode the charges of the fields under $U(1)_t$ and $U(1)_c$. The crosses denote gauge singlets flipping the diagonal mesons. On the r.h.s. the asymmetric $S$-wall.}
\end{figure*}

In this section, we review the properties of the $4d$ $S$-wall, the $FE[USp(2N)]$ theory \footnote{This theory has been first introduced in \cite{Pasquetti:2019hxf}. See also \cite{Hwang:2020wpd,Garozzo:2020pmz,Hwang:2020ddr,Hwang:2021xyw}.}.
The quiver representation of the theory is given on the left of Fig. \ref{fig:fe}.

The IR global symmetry is
\begin{equation}
USp(2N)_x\times USp(2N)_y\times U(1)_t\times U(1)_c\, ,
\end{equation}
with the enhancement $SU(2)^N_y\to USp(2N)_y $ of the symmetries of the saw and where the charges under $U(1)_t$ and $U(1)_c$ are as specified in Fig. \ref{fig:fe}. Notice in particular that the only fields charged under $U(1)_c$ are those forming the saw of the quiver.
To demonstrate our algorithm, we will use the supersymmetric index \cite{Romelsberger:2005eg,Kinney:2005ej,Dolan:2008qi} of this theory, which is a function of the fugacities for these global symmetries so we will denote it by $\mathcal{I}_{FE}^{(N)}(\vec{x};\vec{y};t;c)$. Its explicit definition can be found in eq. (2.17) of \cite{Bottini:2021vms}.

We will also need an asymmetric $S$-wall which is obtained by turning on the superpotential
\begin{eqnarray}
\label{eq:def}
&\delta \mathcal W_\text{def} = \mathrm{Tr}_y \left[\mathsf J \cdot \mathsf{C}\right] , \quad \mathsf J = \frac12 \left(J-J^T\right) ,\nn \\
&J = \mathbb J_2 \otimes \left(\mathbb O_M \oplus \mathbb J_{N-M}\right) , \quad M < N
\end{eqnarray}
where $\mathrm{Tr}_y$ is taken over the emergent $USp (2 N)_y$ symmetry of the theory. $\mathsf C$ is a matrix collecting the mesonic operators constructed from the bifundamental field between the gauge nodes and the fundamental fields in the saw, which is in the antisymmetric representation of $USp (2 N)_y$ \cite{Pasquetti:2019hxf,Hwang:2020wpd}. The antisymmetric matrix $\mathsf J$ is defined in terms of the $K$-dimensional empty matrix $\mathbb O_K$ and the $K$-dimensional Jordan matrix $\mathbb J_K$. This deformation partially breaks $USp(2 N)_y$ to $USp(2 M) \times SU(2)$ and tunes the fugacities of $FE[USp(2N)]$ as ${y_{M+1}=t^{\frac{N-M-1}{2}}v,\cdots,y_N=t^{-\frac{N-M-1}{2}}v}$ for $M < N$. We schematically represent the resulting theory as on the right of Fig. \ref{fig:fe}.

\begin{figure}[b]
\makebox[\linewidth][c]{\includegraphics[]{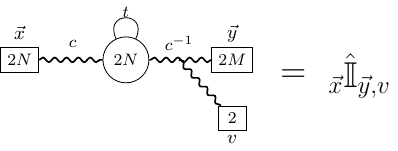} 
}
\caption{\label{identity} Gluing two $S$-walls yields the Identity wall.}
\end{figure}

It was shown in \cite{Bottini:2021vms} that gluing two $S$-walls by gauging a diagonal combination of one $USp(2N)$ from each of them we get the Identity wall, a theory with quantum deformed moduli space whose index behaves as a delta-function that identifies the remaining  symmetries, as shown in Fig. \ref{identity}.
To gauge we add an antisymmetric chiral coupled quadratically to one antisymmetric operator from each block.

At the level of the index the identity property corresponds to
\begin{eqnarray}
\label{eq:delta}
&&{}_{\vec{x}}\hat{\mathbb{I}}_{\vec{y},v}(t)=\oint\udl{\vec{z}_N}\Gd_N(\vec{z};t)\mathcal{I}_{FE}^{(N)}(\vec{x};\vec{z};t;c)\nn\\
&&\mathcal{I}_{FE}^{(N)}(\vec{z};\vec{y},t^{\frac{N-M-1}{2}}v,\cdots,t^{-\frac{N-M-1}{2}}v;t;c^{-1}) \,,
\end{eqnarray}
where we defined the identity operator
\begin{equation}
\label{eq:idop}
\medmath{{}_{\vec{x}}\hat{\mathbb{I}}_{\vec{y},v}(t)=\frac{{\displaystyle\left.\sum_{\sigma \in S_N,\pm} \prod_{i=1}^{N} 2\pi i x_i \delta\left(x_i-y_{\sigma(i)}^{\pm1}\right)\right|_{y_{M+j}=t^{\frac{N-M+1-2j}{2}}v}}}{\Gd_N (\vec x;t) },}
\end{equation}
with the summation $\sum_{\gs\in S_N} \sum_{\pm}$ spanning the Weyl group of $USp(2 N)$ and $j = 1,\dots,N-M$. We also defined with $\udl{\vec{z}_N}$ the $USp(2N)$ integration measure including the Weyl symmetry factor and with $\Gd_N(\vec{z};t)$ the contribution of the $USp(2N)$ vector and antisymmetric chiral multiplets. For their explicit definitions, see eqs. (2.7)-(2.8) of \cite{Bottini:2021vms}.

This duality and various generalisations where the $S$-walls are glued adding some fundamental chirals in the middle $USp(2N)$ gauge node were derived in \cite{Bottini:2021vms} with iterative applications of the IP duality.

\section{Basic duality moves}

We will now introduce the two basic duality moves we will need to perform the local dualisation.
These moves can be considered as the field theory analogue of the local S-action on the  5-branes.

\subsection{Triangle block dualisation}

\begin{figure*}[bt]
\makebox[\linewidth][c]{\includegraphics[]{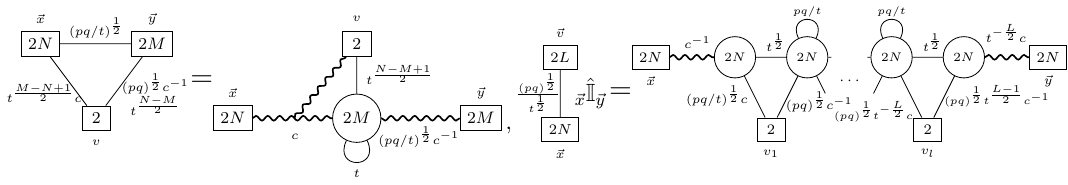} 
}
\caption{\label{basic} On the left, the basic move for the dualisation of a bifundamental block including the $SU(2)_v$ fundamental chirals. On the right, the basic move for the dualisation of a block of $2L$ fundamentals including an Identity wall. For chiral fields the powers of fugacities represent the charges under the corresponding abelian symmetries and the R-charges are encoded in $(pq)^{\frac{R}{2}}$.}
\end{figure*}

The first move replaces a bifundamental block by a fundamental chiral sandwiched between two $S$-walls, as on the left of Fig. \ref{basic}.
This duality  has been derived in \cite{Bottini:2021vms} by iterative use of the IP duality. At the level of the supersymmetric index we have
\begin{eqnarray}
\label{eq:id1}
&&\medmath{\mathcal{I}_{\bigtriangledown}^{(N,M)}\left(\vec{x};\vec{y};v;t;ct^{\frac{M-N}{2}}\right)=\prod_{i=1}^{N-M}\frac{\Gpq{t^{1-i}c^2}}{\Gpq{pq\,t^{-i}}}\oint\udl{\vec{z}_M}\Gd_M\left(\vec{z};t\right)}\nn\\
&&\medmath{\prod_{i=1}^M\Gpq{t^{\frac{N-M+1}{2}}v^{\pm1}z_i^{\pm1}}\mathcal{I}_{FE}^{(M)}\left(\vec{z};\vec{y};t;(pq/t)^{\frac{1}{2}}c^{-1}\right)}\nn\\
&&\medmath{\mathcal{I}_{FE}^{(N)}\left(\vec{x};\vec{z},t^{\frac{N-M-1}{2}}v,\cdots,t^{-\frac{N-M-1}{2}};t;c\right)}\,,
\end{eqnarray}
where we defined the index of the triangle block as
\begin{eqnarray}
\label{eq:tri}
&&\mathcal{I}_{\bigtriangledown}^{(N,M)} \left(\vec x;\vec y;v;t;c\right) = \prod_{i = 1}^N \prod_{j = 1}^M \Gpq{(pq/t)^\frac12 x_i^{\pm1} y_j^{\pm1}} \nn\\
&&\prod_{i = 1}^N \Gpq{t^\frac12 c v^{\pm1} x_i^{\pm1}} \prod_{j = 1}^M \Gpq{(pq)^\frac12 c^{-1} v^{\pm1} y_j^{\pm1}}
\end{eqnarray}
where the definition of the elliptic gamma function $\Gpq{z}$ is given by
\begin{eqnarray}
\Gpq{z}\equiv \Gpq{z;p,q}  = \prod_{n,m=0}^\infty \frac{1-p^{n+1} q^{m+1} z^{-1}}{1- p^{n}  q^{m} z}\,.
\end{eqnarray}

\subsection{Fundamental block dualisation}

The second basic move replaces a block of $2L$ fundamentals times the Identity wall by $L$ triangle blocks sandwiched between two $S$-walls, as on the right of Fig. \ref{basic}.
This can be obtained starting from the duality for the gluing of two $S$-walls with $2L$ chirals in the middle given in \cite{Bottini:2021vms} by gluing two further $S$-walls on each side of the duality. Using the delta property of Fig. \ref{identity} on the l.h.s. of the duality and the flip-flip duality of $FE[USp(2N)]$ \footnote{See eq. (2.14) of \cite{Bottini:2021vms}}  on the r.h.s., we arrive at our basic move. At the level of the supersymmetric index, this reads
\begin{eqnarray}
\label{eq:id2}
&&\medmath{{}_{\vec{x}}\hat{\mathbb{I}}_{\vec{y}}(t)\prod_{i=1}^N\prod_{j=1}^L\Gpq{(pq/t)^{\frac{1}{2}}v_j^{\pm1}x_i^{\pm1}}=\oint\prod_{k=0}^L\udl{w^{(k)}_N} \Gd_N(\vec{w}^{(0)})} \nn\\
&&\medmath{\mathcal{I}_{FE}^{(N)}(\vec{x};\vec{w}^{(0)};t;c^{-1}) \prod_{i=1}^L\mathcal{I}_{\bigtriangledown}^{(N,M)} \left(\vec{w}^{(i-1)};\vec{w}^{(i)};v_i;pq/t;ct^{\frac{1-i}{2}}\right)}\nn\\
&&\medmath{\prod_{k=1}^{L-1}\Gd_N(\vec{w}^{(k)};pq/t) \, \mathcal{I}_{FE}^{(N)}(\vec{w}^{(L)};\vec{y};t;t^{-\frac{L}{2}}c) \, \Gd_N(\vec{w}^{(L)}) \,,}
\end{eqnarray}
where ${}_{\vec{x}}\hat{\mathbb{I}}_{\vec{y}}$ is the identity operator ${}_{\vec{x}}\hat{\mathbb{I}}_{\vec{y},v}$ of \eqref{eq:idop} for $M=N$, in which case it is independent of $v$, and $\Gd_N(\vec{w})$ is the $USp(2 N)$ vector multiplet contribution, whose definition can be found in (2.19) of \cite{Bottini:2021vms}.

\section{Dualisation algorithm}

Given the identity property of the $S$-wall and the basic duality moves, we can use them to derive the $4d$ mirror of any of the $E_\rho^\sigma[USp(2N)]$ theories of \cite{Hwang:2020wpd}. 

The algorithm works as follows:

\begin{enumerate}
\item Chop the quiver by ungauging the gauge nodes into either triangle or fundamental blocks.

\item Dualise each block using the basic duality moves in Fig. \ref{basic}.

\item Glue back the dualised blocks  producing Identity walls to arrive at a quiver with no $S$-walls left. At this stage some operators can acquire a VEV.

\item If some operators acquired a  VEV, follow the RG flow to the IR final configuration, which coincides with the expected mirror of the original theory.
\end{enumerate}

\begin{figure*}[t]
\makebox[\linewidth][c]{\includegraphics[]{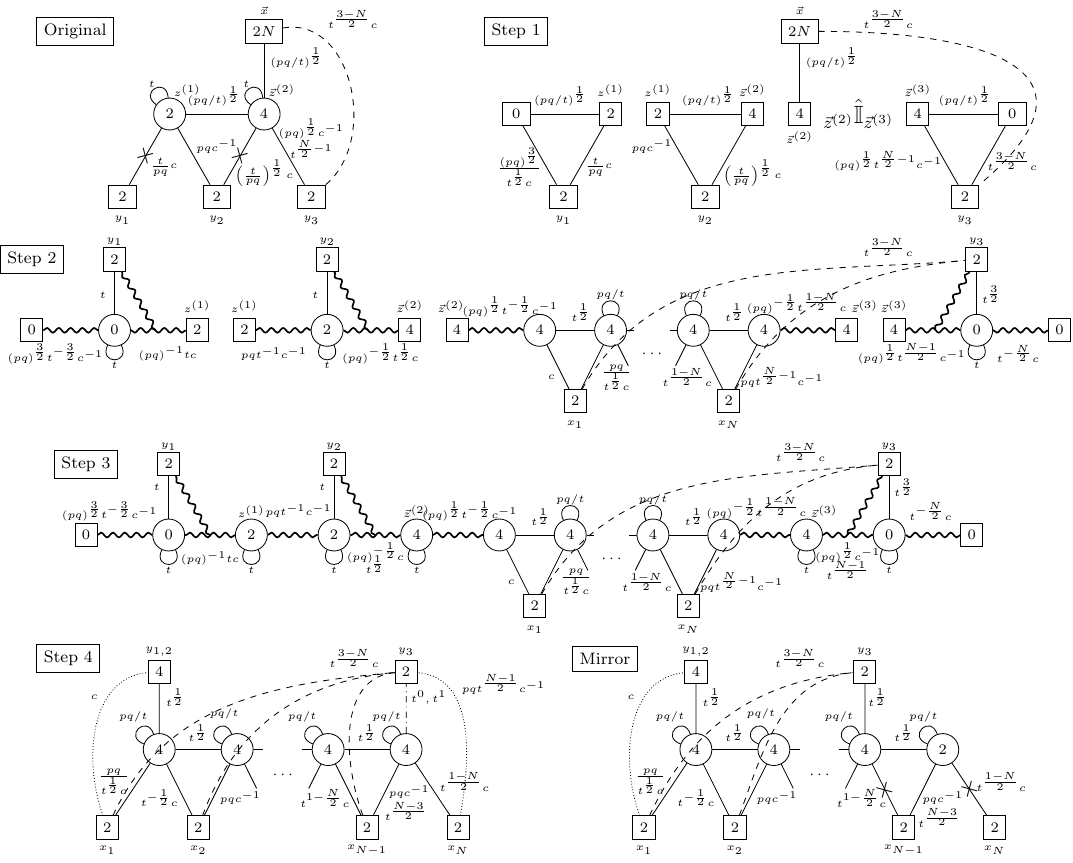} 
}
\caption{\label{fig:prlex} An example of the dualisation algorithm. Dashed lines denote singlets trasnforming under the non-abelian global symmetry that are present from the beginning, while dotted lines denote singlets that are produced during the dualisation. The alternating dashed-dotted line at Step 4 denotes a pair of chirals with charges respectively 0 and 1 under $U(1)_t$ only. We also specify the fugacities of the $i$-th gauge node as $\vec{z}^{(i)}$ to facilitate the comprehension of how the block decomposition is performed. We don't draw singlets uncharged under the non-abelian symmetries except for the initial and final frame, where they are represented by crosses.}
\end{figure*}

Let us exemplify this procedure in the case of $\rho=[N-2,1^2]$ and $\sigma=[1^N]$. This is summarised in Fig. \ref{fig:prlex}.
We start from the quiver of $E_{[N-2,1^2]}[USp(2N)]$ represented on the top left corner of Fig. \ref{fig:prlex}. The crosses represent gauge singlets flipping the corresponding diagonal mesons, while the blue lines denote singlets charged under some of the non-abelian global symmetries. 
For simplicity we omit drawing singlets that don't transform under the non-abelian symmetries in the intermediate steps. One can keep track of them with the index and check that they work out as expected.

In Step 1 we split the quiver into triangle and fundamental blocks. Notice that the fundamental block includes the identity operator.  We can add such operator in the quiver by introducing an auxiliary gauge node labeled by the fugacity $\vec{z}^{(3)}$ in the drawing. We have also completed the first and third triangle adding trivial fields.

In Step 2 we dualise each block using the basic moves. 

In Step 3 we glue back the dualised blocks by restoring the gauging of the original nodes. These three gaugings glue together $S$-walls with the correct charges to yield Identity walls
as in Fig.  \ref{identity}.

In this way we  remove all the $S$-walls from the quiver (the $S$-walls connecting zero nodes are trivial and can be dropped) and we arrive at Step 4, producing also new singlets charged under the non-abelian symmetries which we draw in green. Notice that one set of them gives mass to some of the original blue singlets.

We now have a quiver with no $S$-walls and with fixed charges for the chiral fields. In particular, the orange line denotes a pair of chirals in the bifundamental of the $USp(4)$ gauge and the $SU(2)$ flavor node. One of them has charge 1 under $U(1)_t$ only, while the other is uncharged under every abelian symmetry including the R-symmetry. Such vanishing charges for the latter chiral signal that some operator is acquiring a VEV. Indeed, we note that there is a set of gauge singlets, originating from $\Gd_N(\vec x,t)$ of \eqref{eq:idop}, that couple to the mesons constructed from the bifundamental chirals denoted by the orange line. The superpotential \eqref{eq:def} of the deformed $FE[USp(2N)]$ is yields a  linear superpotential for one of those extra singlets, whose equation of motion  leads to a non-zero VEV of one of the mesons, specifically the one constructed from the bifundamental chirals with no abelian charges.

We can efficiently study the VEV through the supersymmetric index with the technique described in \cite{Gaiotto:2012xa}. Specifically, the index contribution of the aforementioned chirals is $\prod_{i=1}^2\Gamma_e\left(y_3^{\pm1}u_i^{\pm1}\right)$, where $u_i$ are the $USp(4)$ gauge fugacities. From these gamma functions we have two sets of poles that pinch the integration contour at, say, $u_2=y_3^{\pm1}$. 
Taking these residues we obtain the index of the theory after the Higgsing induced by the VEV. In this case, the last $USp(4)$ node is higgsed down to $USp(2)$ and the two chirals  move to the $USp(4)$ node on its left. Hence, we end up with the quiver on the bottom right of Fig. \ref{fig:prlex}, where we are now drawing all the singlets and the charges to show that this indeed coincides with the $E^{[N-2,1^2]}[USp(2N)]$ according to the conventions of \cite{Hwang:2020wpd}. We recovered in this way the mirror duality between $E_{[N-2,1^2]}[USp(2N)]$ and $E^{[N-2,1^2]}[USp(2N)]$.

\section{Comments and outlook}

Our algorithm  dualises the $E_\rho^\sigma[USp(2 N)]$ theory into its mirror dual by acting with two basic duality moves 
and the properties of the $S$-wall. As shown in \cite{Bottini:2021vms}, when we consider gluings involving gauging  manifest symmetries,  everything can be derived  from the Intriligator--Pouliot duality.

However, to actually glue back all the dualised blocks, we need 
the basic moves and the $S$-wall properties with gauging of both manifest and emergent symmetries.
These equivalent relations  can be trivially obtained using the self-duality property of the $FE[USp(2 N)]$ theory,
which follows from the  self-mirror property of $E[USp(2 N)]$. So it would seem that our algorithm to construct mirrors has to assume mirror symmetry at some point.

Nevertheless, to derive the mirror of  $E_\rho^\sigma[USp(2 N)]$, we only need to assume the self-mirror property 
 of $E[USp(2 K)]$ with $K < N$. Since $E[USp(2)]$ is simply a Wess--Zumino model which is manifestly self-mirror, by mathematical induction we can prove that all the mirror dualities of the $E_\rho^\sigma[USp(2 N)]$ family can be derived by the iterative use of the IP duality alone.

One can obtain an analogous algorithm for the local dualisation of $3d$ linear quivers, by
either  taking the $3d$ limit, combined with Coulomb branch VEVs  and real masses, of our $4d$ results
or re-deriving all the basic moves directly in $3d$ by iteritive applications of  the Aharony duality \cite{Aharony:1997gp}, to which the IP duality reduces.

The basic moves in this case can be directly interpreted as the transformations of the NS5 and D5-branes in the brane set-ups of the $3d$ theories under the $S$ element of $SL(2,\mathbb{Z})$.
After dualising the 5-branes in the brane set-up, one usually needs to move D5 across NS5-branes using Hanany--Witten moves to reach a configuration where one can read off the $3d$ gauge theory. Interestingly, in our procedure we don't have to implement the HW moves, but we need to study RG flows initiated by VEVs which
have the effect of moving the matter fields and changing the ranks so to arrive at the final mirror theory.

One should note that our algorithm is not just a prescription to generate integral   identities for the supersymmetric index on $S^3 \times S^1$; while we have provided the supersymmetric index as one concrete example of observables realising our dualisation algorithm, other partition functions can also be manipulated in the same way to derive mirror symmetry from the IP duality.
In fact, as we already emphasised, our algorithm should be regarded as a procedure at the level of field theories.

Indeed the basic duality moves and the identity wall property used in our algorithm can be proven as in \cite{Bottini:2021vms} by iterations of the IP duality, with a procedure which  can be implemented on the UV Lagrangian.

A key ingredient of our algorithm is the possibility gauging emergent symmetries, which  actually allows us to piece-wise dualise and glue black the  triangle or fundamental blocks.
These manipulations are then implemented in the IR.

%One should note that our algorithm is not just a derivation of an identity for the supersymmetric index on $S^3 \times S^1$; while we have provided the supersymmetric index as one concrete example of observables realising our dualisation algorithm, other partition functions can also be manipulated in the same way to derive mirror symmetry from the IP duality.

%In fact, as we already emphasised, our algorithm should be regarded as a procedure at the level of field theories. The basic duality moves and the identity wall property used in our algorithm can be proven as in \cite{Bottini:2021vms} by iterations of the IP duality. Since two $FE[USp(2 N)]$ blocks are glued by gauging manifest flavour symmetries in this derivation, the whole procedure of the derivation can be implemented on the UV Lagrangian with accompanying RG flows understood. Then, to prove mirror symmetry from the basic duality moves, we need another key ingredient of our algorithm: gauging emergent symmetries in the IR, which leads to another version of basic duality moves where the blocks are glued by gauging emergent symmetries instead of the manifest ones. Although such symmetries exist only in a low energy regime, as we already argued, the existence of the symmetries and the subsequent basic duality moves can be proven by mathematical induction without referring to a particular partition function, such as the supersymmetric index we considered in this paper, which indicates that our algorithm is a genuine property of field theories rather than specific partition functions.

Furthermore, as mentioned in the Introduction, our algorithm is the generalisation of the piecewise dualisation of abelian mirror symmetry. For the latter, notably, the same idea has also been used to derive some non-supersymmetric abelian dualities from simpler building blocks \cite{Karch:2016sxi,Seiberg:2016gmd,Karch:2016aux}. We thus expect that our result will provide a new approach to understanding non-abelian dualities with less supersymmetry.

The results of this paper can be generalised in many directions. For example, the same technique can be used to derive the mirror dualities of circular quivers and of the $3d$ $S$-fold SCFTs \cite{Terashima:2011qi,Gang:2015bwa,Gang:2015wya,Assel:2018vtq,Gang:2018huc,Garozzo:2018kra,Garozzo:2019hbf,Garozzo:2019ejm,Beratto:2020qyk} and their $4d$ counterpart.

Moreover,  one can also try to find the basic duality moves corresponding to the local action of other  $SL(2,\mathbb{Z})$ elements,  including the action of the $T$ generator, which in $3d$ corresponds to the introduction of a Chern--Simons
coupling. This will allow us to  generate  more general pairs of $4d$ dual theories. We plan to investigate this in a future work.

\begin{acknowledgments}
We would like to thank  L.~E.~Bottini and D.~Zhang for useful discussions. C.H. is partially supported by the STFC consolidated grant ST/T000694/1. S.P. and M.S. are partially supported by the  INFN.   M.S. is also partially supported by the University of Milano-Bicocca grant 2016-ATESP0586 and by the MIUR-PRIN contract 2017CC72MK003.
\end{acknowledgments}

\bibliography{ref}

\end{document}